\documentclass[12pt]{article}
 \usepackage{epsfig}
 \def\be{\begin{equation}}
 \def\ee{\end{equation}}
 \def\bea{\begin{eqnarray}}
 \def\eea{\end{eqnarray}}
 \usepackage{graphicx}

 \catcode`\@=11
 \def\lsim{\mathrel{\mathpalette\@versim<}}
 \def\gsim{\mathrel{\mathpalette\@versim>}}
 \def\@versim#1#2{\vcenter{\offinterlineskip
 \ialign{$\m@th#1\hfil##\hfil$\crcr#2\crcr\sim\crcr } }}
 \catcode`\@=12

 \catcode`@=12
\begin{document}
 \thispagestyle{empty}
 \begin{flushright}
 UCRHEP-T612\\
 Mar 2021\
 \end{flushright}
 \vspace{1.0in}
 \begin{center}
 {\LARGE \bf Naturally Light Dirac Neutrinos from $SO(10) \times U(1)_\psi$\\}
 \vspace{1.2in}
 {\bf Ernest Ma\\}
 \vspace{0.2in}
{\sl Department of Physics and Astronomy,\\ 
University of California, Riverside, California 92521, USA\\}
\end{center}
 \vspace{1.8in}

\begin{abstract}\
A new solution is presented where the right-handed neutrino 
$\nu_R$ in $SO(10)$ pairs up with $\nu_L$ to form a naturally light Dirac 
neutrino.  It is based on the framework of $E_6 \to SO(10) \times U(1)_\psi$, 
then $SO(10) \to SU(5) \times U(1)_\chi$.  
\end{abstract}

 \newpage
 \baselineskip 24pt

\noindent \underline{\it Introduction}~:~
Neutrinos ($\nu$) are observed but not understood.  They are not massless 
but very light~\cite{pdg20}.  They may be self-conjugate two-component 
spinors (Majorana) or four-component spinors (Dirac) with right-handed 
components ($\nu_R$) which have no electroweak interactions.  Neutrinoless 
double beta decay experiments have so far no conclusive evidence for them 
to be Majorana, which is the prevalent theoretical thinking.

Why is the Dirac option disfavored?  There are two well-known answers. 
(1) To have a Dirac neutrino, $\nu_R$ must exist.  However, under the 
standard-model (SM) gauge symmetry $SU(3)_C \times SU(2)_L \times U(1)_Y$, 
it is a trivial singlet and not mandatory.  If it is added anyway, then it 
is allowed a Majorana mass.  Together with the Dirac mass linking $\nu_L$ 
to $\nu_R$ through the one SM Higgs doublet $\Phi = (\phi^+,\phi^0)$, this 
forms the well-known $2 \times 2$ mass matrix
\begin{equation}
{\cal M}_\nu = \pmatrix{0 & m_D \cr m_D & m_R}.
\end{equation}
With the usual reasonable assumption $m_D << m_R$, the famous seesaw 
mechanism yields a very small Majorana neutrino mass $m_\nu \simeq m_D^2/m_R$. 
(2) To protect $\nu_R$ from having a Majorana mass, a symmetry has to be 
imposed.  The obvious one is lepton number.  In that case, $m_R$ is 
forbidden, but $m_D$ is allowed.  However there is no 
understanding why the Yukawa coupling which generates $m_D$ is so small,  
$10^{-11}$ or less, since $\langle \phi^0 \rangle = 174$ GeV and 
$m_\nu < 1.1$ eV~\cite{katrin19}.
To have a compelling case for Dirac neutrinos, two conditions must be 
met. 
\begin{itemize}
\item (A) The existence of $\nu_R$ should not be {\it ad hoc}, but based on 
a well-motivated theoretical framework, within which it should not acquire 
a Majorana mass. 
\item (B) The smallness of the Dirac neutrino mass should be obtained 
naturally, without any extra symmetry. 
\end{itemize}
In past studies~\cite{mp17}, limited success has been achieved, but at the 
expense of extra symmetries, some of which are softly broken.  Here a new 
solution based on $E_6 \to SO(10) \times U(1)_\psi$, then 
$SO(10) \to SU(5) \times U(1)_\chi$ is presented which satisfies for the 
first time both (A) and (B) in full measure.

\noindent \underline{\it Essential Existence of $\nu_R$}~:~
To justify the presence of $\nu_R$, it ought to transform under a symmetry 
related to those of the SM.  It could be a global symmetry such as lepton 
number mentioned earlier.  However, it would be more convincing and 
compelling if it were a gauge symmetry such as $U(1)_{B-L}$~\cite{d79} 
or $SU(2)_R \times U(1)_{B-L}$~\cite{mm80}.  A recently proposed alternative 
is a gauge $U(1)_D$ symmetry~\cite{m21,m21-1} not related to the SM but 
essential for dark matter, with $\nu_R$ as the bridge between the two sectors.

The breaking of $U(1)_{B-L}$ [and $SU(2)_R$] is usually assumed without 
hesitation to allow $\nu_R$ to obtain a large Majorana mass, thereby 
already not satisfying condition (A).  To avoid this eventuality, there 
is a simple solution.  This breaking does not have to be 
$\Delta L = 2$.  If it is $\Delta L = 3$ for example, then neutrinos 
are Dirac.  This was first pointed out~\cite{mpr13} for a general $U(1)_X$ 
symmetry and applied~\cite{ms15} to $U(1)_L$ for Dirac neutrinos.
However, this mechanism does not by itself explain why the neutrino 
Higgs Yukawa couplings are so small.

The existence of $\nu_R$ may also be justified in $SU(6)$ as shown 
recently~\cite{m21-2,m21-3}.  However, it is best known as the 
missing link which allows the 15 fermions (per family) of the SM to 
form a complete $\underline{16}$ representation of $SO(10)$.  Whereas the 
usual study of $SO(10)$ proceeds from its left-right decomposition 
$SU(3)_C \times SU(2)_L \times SU(2)_R \times U(1)_{B-L}$, here 
the $SU(5) \times U(1)_\chi$ alternative is considered: 
\begin{equation}
\underline{16} = (5^*,3) + (10,-1) + (1,-5),
\end{equation}
where
\begin{equation}
(5^*,3)  = \pmatrix {d^c \cr d^c \cr d^c \cr e \cr \nu}, ~~~ 
(10,-1) = \pmatrix {0 & u^c & u^c & -u & -d \cr -u^c & 0 & u^c & -u & -d 
\cr  u^c & -u^c & 0 & -u & -d \cr u & u & u & 0 & -e^c  \cr 
d & d & d & e^c & 0}, ~~~ (1,-5) = \nu^c.
\end{equation}
It has been shown~\cite{m18} that the $U(1)_\chi$ charge may be used as a 
marker for dark matter, with dark parity given by $(-1)^{Q_\chi+2j}$. 
Here it will be shown how a naturally small Dirac mass linking $\nu$ to 
$\nu^c$ is obtained instead.

\noindent \underline{\it Naturally Small Dirac Neutrino Mass}~:~
To obtain a naturally small Dirac neutrino mass, the mechanism of 
Ref.~\cite{m01} is the simplest solution.  Let there be  
two Higgs doublets, say $\Phi = (\phi^+,\phi^0)$ and 
$\eta = (\eta^+,\eta^0)$ which are distinguished by some symmetry, 
so that $\bar{\nu}_R \nu_L$ couples to $\eta^0$, but not $\phi^0$.  
This symmetry is then broken by the soft dimension-two 
$\mu^2 \Phi^\dagger \eta$ term, with $m^2_\Phi < 0$ as usual, but  
$m^2_\eta >0$ and large.  Consequently, the vacuum expectation value  
$\langle \eta^0 \rangle$ is given by $-\mu^2 \langle \phi^0 \rangle/m^2_\eta$, 
which is naturally small, implying thus a very small Dirac neutrino mass. 
In the original application~\cite{m01}, $\nu_R$ also has a large 
Majorana mass, hence the $\nu_L$ mass is doubly suppressed.  In that 
case, $m_\eta$ could well be of order 1 TeV.  On the other hand, if the 
symmetry and the particle content are such that $\nu_R$ is prevented 
from having a Majorana mass, then a much larger $m_\eta$ works just as 
well for a tiny Dirac neutrino mass.

Recently this mechanism has been used~\cite{m21,m21-1} in the framework of 
a gauge $U(1)_D$ symmetry under which the SM particles do not transform, 
but $\nu_R$ and other fermion singlets do.  The $U(1)_D$ symmetry is broken 
by singlet scalars which transform only under $U(1)_D$.  The connection 
between the SM and this new sector is a set of Higgs doublets which transform 
under both, so that $\nu_R$ pairs up with $\nu_L$ to form a Dirac neutrino.  
The particle content is chosen such that global lepton number is 
conserved as well as a dark parity or dark number.

A more compelling case~\cite{m21-3} is to identify $\nu_R$ as part of the 
fundamental representation of $SU(6)$ which breaks to $SU(5) \times U(1)_N$.  
Here a new solution is presented with $E_6 \to SO(10) \times U(1)_\psi$, 
then $SO(10) \to SU(5) \times U(1)_\chi$.

\noindent \underline{\it Sources of Quark and Lepton Masses}~:~
In the SM, there is only one Higgs doublet.  All quark and lepton masses 
come from just this one source.  In the conventional left-right model, 
where $\nu_R$ must appear as part of an $SU(2)_R$ doublet, the common 
source of all quark and lepton masses is a Higgs $SU(2)_L \times SU(2)_R$ 
scalar bidoublet.  It is thus not possible to separate out the Dirac 
neutrino mass from the others.  In the case of $SO(10)$ which encompasses 
the left-right symmetry, consider how the fermions of the $\underline{16}$ 
acquire mass.  They do so from coupling to the scalars
\begin{eqnarray}
\underline{10} &=& (5,2) + (5^*,-2), \\ 
\underline{120} &=& (5,2) + (5^*,-2) + (10,-6) + (10^*,6) + (45,2) + 
(45^*,-2), \\ 
\underline{126}^* &=& (1,10) + (5,2) + (10^*,6) + (15^*,-6) + (45^*,-2) + 
(50^*,2),  
\end{eqnarray}
under $SU(5) \times U(1)_\chi$.

From the $\underline{16}_F \times \underline{16}_F \times \underline{10}_S$ 
Yukawa term, it is seen that $d^c d$ and $e e^c$ couple to $(5^*,-2)$, 
whereas $u^c u$ and $\nu \nu^c$ couple to $(5,2)$.  All would acquire 
masses if the scalar doublets
\begin{equation} 
\Phi_1 = (\phi_1^0,\phi_1^-) \sim (1,2,-1/2,-2), ~~~  
\Phi_2 = (\phi_2^+,\phi_2^0) \sim (1,2,1/2,2)
\end{equation} 
under $SU(3)_C \times SU(2)_L \times U(1)_Y \times U(1)_N$ develop nonzero 
vacuum expectation values.  However, 
\begin{equation}
\tilde{\Phi}_1 = i \sigma_2 \Phi_1^* = (\phi_1^+,-\bar{\phi}_1^0), ~~~ 
\tilde{\Phi}_2 = i \sigma_2 \Phi_2^* = (\bar{\phi}_2^0,-\phi_2^-)
\end{equation}
transform exactly as $\Phi_{2,1}$ respectively.  Hence both $\Phi_{1,2}$ 
would couple to all quarks and leptons.  This is the analog of the well-known 
property of the $SU(2)_L \times SU(2)_R$ scalar bidoublet, i.e.
\begin{equation}
\eta = \pmatrix {\eta_1^0 & \eta_2^+ \cr \eta_1^- & \eta_2^0},
\end{equation}
where
\begin{equation}
\tilde{\eta} = \sigma_2 \eta^* \sigma_2 = \pmatrix {\bar{\eta}_2^0 & 
-\eta_1^+ \cr -\eta_2^- & \bar{\eta}_1^0},
\end{equation}
which transforms exactly as $\eta$.

To distinguish $\Phi_{1,2}$ from $\tilde{\Phi}_{2,1}$, the origin of 
the fermion $\underline{16}_F$ of Eq.~(2) and the scalars of Eqs.~(4),(5),(6) 
is assumed to be from $E_6 \to SO(10) \times U(1)_\psi$.  Using
\begin{equation}
\underline{27} = (\underline{16},1) + (\underline{10},-2) + 
(\underline{1},4),  
\end{equation}
the scalars of Eqs.~(4),(5),(6) must then have $Q_\psi = -2$.  Thus 
$\Phi_{1,2}$ transform differently from $\tilde{\Phi}_{2,1}$ which have 
$Q_\psi = 2$.  It is now possible to have two different $(\underline{10},-2)$ 
scalars, one with negative mass-squared which breaks along the $(5^*,-2,-2)$ 
component, so it gives mass to $d^c d$ and $e e^c$, and the other with large 
mass-squared which has an induced tiny vacuum expectation value along the 
$(5,2,-2)$ component and gives mass to $\nu \nu^c$. 

The next step is to find how the $u^c u$ term may be distinguished from 
$\nu \nu^c$.  Under $SU(5) \times U(1)_\chi$, the former comes from 
$(10,-1) \times (10,-1)$ which couples to the scalars $(5,2)$ and 
$(45,2)$, whereas the latter comes from $(5^*,3) \times (1,-5)$ which 
couples only to $(5,2)$.  Now it is well-known that $(45,2)$ also contains 
the doublet $(1,2,1/2,2)$. Let it be called $\Phi_4$.  So if 
$\langle \phi_4^0 \rangle$ is large and $\langle \phi_2^0 \rangle$ is 
small, then the Dirac neutrino mass is guaranteed to be small.  This is 
the key observation of the present study.

\noindent \underline{\it Symmetry Breaking Details}~:~
The relevant scalars of this model are listed in Table~1.
\begin{table}[tbh]
\centering
\begin{tabular}{|c|c|c|c|c|}
\hline
scalar & $SU(2)_L \times U(1)_Y$ & $SU(5) \times U(1)_\chi$ & 
$SO(10) \times U(1)_\psi$ & $E_6$  \\
\hline
$\zeta_0$ & $(1,0)$ & $(1,0)$ & $(\underline{1},0)$ & $\underline{78}$ \\ 
$\zeta_1$ & $(1,0)$ & $(1,0)$ & $(\underline{1},-8)$ & $\underline{351}'$ \\ 
$\zeta_2$ & $(1,0)$ & $(24,0)$ & $(\underline{1050},0)$ & $\underline{5824}^*$ 
\\ 
$\zeta_3$ & $(1,0)$ & $(1,0)$ & $(\underline{45},4)$ & $\underline{351}$ \\ 
$\zeta_4$ & $(1,0)$ & $(1,-5)$ & $(\underline{16},-3)$ & $\underline{2430}$ \\ 
\hline
$\Phi_1$ & $(2,-1/2)$ & $(5^*,-2)$ & $(\underline{10},-2)$ & 
$\underline{27}$  \\ 
$\Phi_2$ & $(2,1/2)$ & $(5,2)$ & $(\underline{10},-2)$ & 
$\underline{27}$  \\ 
$\Phi_3$ & $(2,-1/2)$ & $(45^*,-2)$ & $(\underline{120},-2)$ & 
$\underline{351}$  \\ 
$\Phi_4$ & $(2,1/2)$ & $(45,2)$ & $(\underline{120},-2)$ & 
$\underline{351}$  \\ 
\hline
$\Phi_5$ & $(2,-1/2)$ & $(5^*,-2)$ & $(\underline{126},-2)$ & 
$\underline{1728}$  \\ 
\hline
\end{tabular}
\caption{Scalars for $E_6 \to SO(10) \times U(1)_\psi$, 
$SO(10) \to SU(5) \times U(1)_\chi$, and 
$SU(5) \to SU(3)_C \times SU(2)_L \times U(1)_Y$.}
\end{table}
The scalar $\zeta_0$ breaks $E_6$ to $SO(10) \times U(1)_\psi$; $\zeta_1$ 
breaks $U(1)_\psi$;  $\zeta_2$ breaks $SO(10)$ to 
$SU(3)_C \times SU(2)_L \times U(1)_Y \times U(1)_\chi$.  The corresponding 
vacuum expectation values (VEV) $v_{0,1,2}$ are all big.  The scalar 
$\zeta_3$ breaks $SO(10) \times U(1)_\psi$ to 
$SU(5) \times U(1)_\chi$, but $v_3$ may be small because both $U(1)_\psi$ and 
$SO(10)$ have already been broken by $v_{1,2}$.  Finally, $\zeta_4$ breaks 
$U(1)_\chi$, presumably at a lower scale.  The doublets $\Phi_{1,2,3,4,5}$ 
all break $SU(2)_L \times U(1)_Y$ to $U(1)_Q$ as in the SM.
There are of course other possible scalar doublets, but they are all 
assumed heavy enough, not to affect the outcome of the SM.

In the $\underline{27} \times \underline{27}^* \times \underline{78}$ 
trilinear scalar coupling, $v_0$ splits the $(\underline{10},-2)$ component  
from $(\underline{1},4)$ and $(\underline{16},1)$, but the two $SU(5)$ 
components of $(\underline{10},-2)$, i.e. $\Phi_{1,2}$, are degenerate.  To 
split them with $\Phi_1(\Phi_2)$ having negative (positive) mass-squared, 
the $U(1)_\psi$ charge plays an important role. Whereas 
$(\underline{126}^*,-2)$ comes from $\underline{351}'$, 
$(\underline{126},-2)$ comes from $\underline{1728}$.  They may have 
different masses.  This would not be possible in $SO(10)$ alone. 
Note also that $\Phi_5$ cannot couple to the SM fermions because it 
comes from $\underline{126}$ not $\underline{126}^*$.  The choice of 
$(\underline{1050},0)$ for $\zeta_2$ is because it couples to 
$\underline{10} \times \underline{126}^*$ and 
$\underline{120} \times \underline{126}^*$, but not
$\underline{10} \times \underline{120}$.  This allows $\Phi_2$ not to 
mix with $\Phi_4$, whereas $\Phi_{1,3,5}$ may mix in a $3 \times 3$ 
mass-squared matrix as shown below.

The trilinear scalar $\mu_{15} \Phi_1 \Phi_5^\dagger \zeta_2$ and 
$\mu_{35} \Phi_3 \Phi_5^\dagger \zeta_2$ couplings are allowed by 
$SU(5)$ ($5^* \times 5 \times 24$), ($45^* \times 5 \times 24$), $SO(10)$ 
($\underline{10} \times \underline{126}^* \times \underline{1050}$), 
($\underline{120} \times \underline{126}^* \times \underline{1050}$), and 
$E_6$ ($\underline{27} \times \underline{1728}^* \times \underline{5824}^*$), 
($\underline{351} \times \underline{1728}^* \times \underline{5824}^*$). 
The $\Phi_1 \Phi_3^\dagger \zeta_2$ coupling is forbidden because 
$\underline{10} \times \underline{120} \times \underline{1050}$ is not 
allowed under $SO(10)$.  The resulting $3 \times 3$ mass-squared matrix 
for $\Phi_{1,3,5}$ is given by
\begin{equation}
{\cal M}^2_{1,3,5} = \pmatrix{ m^2_{10} & 0 & \mu_{15} v_2 \cr 0 & m^2_{120} & 
\mu_{35} v_2 \cr \mu_{15} v_2 & \mu_{35} v_2 & m^2_{126}}.
\end{equation}
The corresponding $2 \times 2$ mass-squared matrix for $\Phi_{2,4}$ is
simply
\begin{equation}
{\cal M}^2_{2,4} = \pmatrix{ m^2_{10} & 0 \cr 0 & m^2_{120}}.
\end{equation}
It is thus possible to keep $m^2_{10} > 0$ and $m^2_{120} < 0$, and find 
two negative eigenvalues in ${\cal M}^2_{1,3,5}$, with eigenstates of 
linear combinations of $\Phi_{1,3,5}$.  They will contribute to the 
$d^c d$ and $e e^c$ masses, whereas $\Phi_4$ is responsible for the $u^c u$ 
masses, and since $\Phi_2$ has no VEV at this stage, neutrino masses are 
zero.

The $\Phi_{2,4}$ scalars are possibly connected to $\Phi_{1,3,5}$ through 
$\zeta_3$. The trilinear $\mu_{21} \Phi_2 \Phi_1 \zeta_3$ coupling is 
allowed by $SU(5)$ ($5 \times 5^* \times 1$), $SO(10)$ 
($\underline{10} \times \underline{10} \times \underline{45}$), and $E_6$ 
($\underline{27} \times \underline{27} \times \underline{351}$).  This 
results in 
\begin{equation}
\langle \phi_2^0 \rangle \simeq - {\mu_{21} \langle \phi_1^0 \rangle v_3 
\over m^2_{10}},
\end{equation}
as discussed earlier.  This suppressed VEV is the source of the Dirac 
neutrino mass. As for the $\Phi_2 \Phi_3$ and $\Phi_2 \Phi_5$ terms, they 
are not allowed because the former does not obey $SU(5)$ and the latter 
$U(1)_\psi$.  The only other allowed coupling is $\Phi_3 \Phi_4 \zeta_3$, 
which is expected because it comes from 
$\underline{351} \times \underline{351} \times \underline{351}$. 
Finally $\zeta_4$ transforms as $(1,1,0,-5,-3)$ under 
$SU(3)_C \times SU(2)_L \times U(1)_Y \times U(1)_\chi \times U(1)_\psi$, 
whereas a Majorana neutrino mass for $\nu^c$ would require a singlet 
transforming as $(1,1,0,10,-2)$.  The absence of the latter with a VEV 
means that lepton number remains a global symmetry in the SM.

\noindent \underline{\it Concluding Remarks}~:~
In the context of $E_6 \to SO(10) \times U(1)_\psi$, then 
$SO(10) \to SU(5) \times U(1)_\chi$, it is shown how the neutrino 
obtains a naturally small Dirac mass.  This involves 
five scalar doublets: $\Phi_1, \Phi_3, \Phi_5 \sim (1,2,-1/2,-2)$ 
from the $(5^*,-2)$ of $\underline{10}$ in $SO(10)$, the $(45^*,-2)$ of 
$\underline{120}$, and the $(5^*,-2)$ of $\underline{126}$, all of which 
have $Q_\psi = -2$ under $E_6 \to SO(10) \times U(1)_\psi$; and 
$\Phi_2, \Phi_4, \sim (1,2,1/2,-2)$ from the $(5,2)$ of 
$\underline{10}$ in $SO(10)$, and the $(45,2)$ of $\underline{120}$.  
The key is that whereas $\Phi_{1,2}$ and $\Phi_{3,4}$ are not split by 
themselves, the conjugate of $\Phi_5$ lies in a different $E_6$ multiplet, 
so the would-be $\Phi_6$ may be assumed much heavier and be ignored. 

With the addition of the five singlet scalars of Table~1, it is shown how 
$\Phi_{1,3,4}$ may develop VEV for the $d^c d$, $e e^c$, and $u^c u$ masses.  
The $\nu \nu^c$ masses come only from $\Phi_2$, which 
keeps a large positive mass-squared, so its VEV is induced 
and suppressed by the mechanism of Ref.~\cite{m01}.  The breaking of 
$U(1)_\chi$ comes from $\zeta_4$ which allows global lepton number to remain.

The fermions of this model are the same as in the SM, except that the 
neutrino is Dirac.  They form complete $\underline{16}$ representations 
of $SO(10)$ which may have a remnant $U(1)_\chi$ symmetry~\cite{m18} from 
breaking to $SU(5) \times U(1)_\chi$, instead through the usual left-right 
intermediate step.  There are possibly three scalar 
doublets $(\Phi_{1,3,4})$ at the electroweak scale, together with a 
singlet $\zeta_4$ at the $U(1)_\chi$ breaking scale.  The $Z_\chi$ gauge 
should also appear at this scale, with known couplings to all SM 
particles.  From present collider data, its mass is greater than a few TeV.  

\noindent \underline{\it Acknowledgement}~:~
This work was supported in part by the U.~S.~Department of Energy Grant 
No. DE-SC0008541.

\baselineskip 16pt

\bibliographystyle{unsrt}

\end{document}